\documentstyle{mn}
\newif\ifAMStwofonts
\ifoldfss
  \ifCUPmtlplainloaded \else
    \NewTextAlphabet{textbfit} {cmbxti10} {}
    \NewTextAlphabet{textbfss} {cmssbx10} {}
    \NewMathAlphabet{mathbfit} {cmbxti10} {} % for math mode
    \NewMathAlphabet{mathbfss} {cmssbx10} {} %  "   "    "
  \fi
  \ifAMStwofonts
    \ifCUPmtlplainloaded \else
      \NewSymbolFont{upmath} {eurm10}
      \NewSymbolFont{AMSa} {msam10}
      \NewMathSymbol{\upi}     {0}{upmath}{19}
      \NewMathSymbol{\umu}     {0}{upmath}{16}
      \NewMathSymbol{\upartial}{0}{upmath}{40}
      \NewMathSymbol{\leqslant}{3}{AMSa}{36}
      \NewMathSymbol{\geqslant}{3}{AMSa}{3E}

    \fi
  \fi
\fi % End of OFSS

\ifnfssone
  \newmathalphabet{\mathit}
  \addtoversion{normal}{\mathit}{cmr}{m}{it}
  \addtoversion{bold}{\mathit}{cmr}{bx}{it}
  \newmathalphabet{\mathbfit} % math mode version of \textbfit{..}
  \addtoversion{normal}{\mathbfit}{cmr}{bx}{it}
  \addtoversion{bold}{\mathbfit}{cmr}{bx}{it}
  \newmathalphabet{\mathbfss} % math mode version of \textbfss{..}
  \addtoversion{normal}{\mathbfss}{cmss}{bx}{n}
  \addtoversion{bold}{\mathbfss}{cmss}{bx}{n}
  \ifAMStwofonts
    \ifCUPmtlplainloaded \else
      \UseAMStwoboldmath
      \makeatletter
      \new@mathgroup\upmath@group
      \define@mathgroup\mv@normal\upmath@group{eur}{m}{n}
      \define@mathgroup\mv@bold\upmath@group{eur}{b}{n}
      \edef\UPM{\hexnumber\upmath@group}
      \new@mathgroup\amsa@group
      \define@mathgroup\mv@normal\amsa@group{msa}{m}{n}
      \define@mathgroup\mv@bold\amsa@group{msa}{m}{n}
      \edef\AMSa{\hexnumber\amsa@group}
      \makeatother
      \mathchardef\upi="0\UPM19
      \mathchardef\umu="0\UPM16
      \mathchardef\upartial="0\UPM40
      \mathchardef\leqslant="3\AMSa36
      \mathchardef\geqslant="3\AMSa3E
    \fi
  \fi
\fi % End of NFSS release 1

\ifnfsstwo
  \DeclareMathAlphabet{\mathbfit}{OT1}{cmr}{bx}{it}
  \SetMathAlphabet\mathbfit{bold}{OT1}{cmr}{bx}{it}
  \DeclareMathAlphabet{\mathbfss}{OT1}{cmss}{bx}{n}
  \SetMathAlphabet\mathbfss{bold}{OT1}{cmss}{bx}{n}
  \ifAMStwofonts
    \ifCUPmtlplainloaded \else
      \DeclareSymbolFont{UPM}{U}{eur}{m}{n}
      \SetSymbolFont{UPM}{bold}{U}{eur}{b}{n}
      \DeclareSymbolFont{AMSa}{U}{msa}{m}{n}
      \DeclareMathSymbol{\upi}{0}{UPM}{"19}
      \DeclareMathSymbol{\umu}{0}{UPM}{"16}
      \DeclareMathSymbol{\upartial}{0}{UPM}{"40}
      \DeclareMathSymbol{\leqslant}{3}{AMSa}{"36}
      \DeclareMathSymbol{\geqslant}{3}{AMSa}{"3E}
    \fi
  \fi
\fi % End of NFSS release 2

\ifCUPmtlplainloaded \else
  \ifAMStwofonts \else % If no AMS fonts
    \def\upi{\pi}
    \def\umu{\mu}
    \def\upartial{\partial}
  \fi
\fi

\title{Cosmic Equation of state from Strong Gravitational Lensing
Systems}

\author[M. Biesiada,  A. Pi{\'o}rkowska \& B. Malec]
  {Marek Biesiada \thanks{biesiada@us.edu.pl},
  Aleksandra Pi{\'o}rkowska  \thanks{apiorko@us.edu.pl}  and Beata Malec \thanks{beata.malec@us.edu.pl}\\
  Department of Astrophysics and Cosmology, Institute of Physics, University of
  Silesia, Uniwersytecka 4, 40-007 Katowice, Poland }

\pagerange{\pageref{firstpage}--\pageref{lastpage}} \pubyear{2010}

\begin{document}

\maketitle

\label{firstpage}

\begin{abstract}

Accelerating expansion of the Universe is a great challenge for
both physics and cosmology. In light of lacking the convincing
theoretical explanation, an effective description of this
phenomenon in terms of cosmic equation of state turns out useful.

The strength of modern cosmology lies in consistency across
independent, often unrelated pieces of evidence. Therefore, every
alternative method of restricting cosmic equation of state is
important. Strongly gravitationally lensed quasar-galaxy systems
create such new opportunity by combining stellar kinematics
(central velocity dispersion measurements) with lensing geometry
(Einstein radius determination form position of images).

In this paper we apply such method to a combined data sets from
SLACS and LSD surveys of gravitational lenses. In result we obtain
the cosmic equation of state parameters, which generally agree
with results already known in the literature. This demonstrates
that the method can be further used on larger samples obtained in
the future. Independently noticed systematic deviation between
fits done on standard candles and standard rulers is revealed in
our findings. We also identify an important selection effect
crucial to our method associated with geometric configuration of
the lensing system along line of sight, which may have
consequences for sample construction from the future lensing
surveys.

\end{abstract}


\section{Introduction}

The present acceleration of the cosmic expansion is a fundamental
challenge to standard models of both particle physics and
cosmology. Discovery of this phenomenon on the Hubble diagrams
obtained from the SNIa surveys (Riess et al. 1998, Perlmutter et
al. 1999,Riess et al. 2004,Wood-Vasey et al. 2007; Davis et al.
2007; Kowalski et al.2008 ) in combination with independent
estimates of the amount of baryons and cold dark matter (Spergel
et al. 2003, Eisenstein et al. 2005) led us to believe that most
of the energy in the Universe exists in the form of mysterious
dark energy.

The new physics of dark energy may lie in the nature of gravity,
the quantum vacuum, or extra dimensions. Concerning the first
possibility there exists an increasing body of literature (e.g.
Buchert 2001, R{\"a}s{\"a}nen 2004, Ellis \& Buchert 2005,
Wiltshire 2007) pointing out that if one attempts to average out
local sources of gravity (galaxies and clusters) in order to
obtain the smoothed description of the Universe in the large such
averaging procedure – not commuting with temporal evolution –
could manifest as an additional source term in the energy-momentum
tensor. Within the second possibility our ideas about the quantum
vacuum are expressed by either introducing cosmological constant
$\Lambda$ or some time evolving scalar field (quintessence). The
last possibility is to contemplate modifications to the
Friedman-Robertson-Walker models arising in brane-world scenarios.
Irrespective of the theoretical approach chosen a common point
with the observations usually occurs at the level of the $w(z)$
coefficient in an effective equation of state $p=w(z) \rho$ for
dark energy.

The potential of constraining dark energy models with SNIa data,
even though ever increasing, would not be sufficient if taken
alone in separation form the other approaches. Indeed, the power
of modern cosmology lies in building up consistency rather than in
single, precise, crucial experiments. Therefore, every alternative
method of restricting cosmological parameters is desired. In this
spirit a number of combined analyses involving lensing statistics
(Silva \& Bertolami 2003), CMBR measurements (Spergel et al. 2003,
Hinshaw et al. 2009), age-redshift relation (Alcaniz, Jain \& Dev
2003), x-ray luminosities of galaxy clusters (Allen et al. 2008)
or the large scale structure considerations (Eisenstein et al.
2005) have been performed in the literature (references above
being far from complete).

In this paper we use strongly gravitationally lensed systems for
providing additional constraints on dark energy models.  The idea
of using such systems for measuring the cosmic equation of state
 was discussed in Biesiada (2006)
and also in more recent paper by Grillo et al. (2008). The first
(to our knowledge) formulations of this approach can be traced
back to Futamase \& Yoshida (2001). Next sections outline the
method, the sample used and cosmological scenarios tested. The
last section presents the results and conclusions.

\section{The Method}

Strong gravitational lensing occurs whenever the source, the lens
and observer are so well aligned that the observer--source
direction lies inside the so called Einstein ring %(more precisely -- the critical curve)
of the lens. In cosmological context the source is usually a
quasar with a galaxy acting as the lens. Although strong lensing
by
 clusters is known and number of such cases increases, we will be
concerned with galaxies acting as lenses. For detailed theory of
gravitational lensing see e.g. Schneider, Ehlers \& Falco (1992).
Strong lensing reveals itself as multiple images of the source.
The image separations in the system depend on angular-diameter
distances to the lens and to the source, which in turn are
determined by background cosmology. This opens a possibility to
constraining the cosmological model provided that we have good
knowledge of the lens model.

Since the discovery of the first gravitational lens the number of
strongly lensed systems increased to a hundred (in the CASTLES
database \footnote{http://www.cfa.harvard.edu/castles/}) and is
steadily increasing following the new surveys like SLACS (Sloan
Lens ACS Survey \footnote{http://www.slacs.org/}). It turns out
that in vast majority of cases the lens is a late type E/SO
galaxy. This could be understood since ellipticals being a
latecomers in hierarchical structure formation are created in
mergers of low-mass spiral galaxies. Hence they are more massive
than spirals and because the Einstein ring radius scales with
mass, the probability of their acting as lenses is higher. Despite
they lack bright kinematic tracers at large radii (e.g. like HI in
disk galaxies) and thus their kinematics is more difficult to
measure, there exists an increasing evidence that their mass
density profile can well be approximated by singular isothermal
sphere (SIS) model (or a variant thereof called singular
isothermal ellipsoid -- SIE).

Now, the idea is that formula for the Einstein radius in a SIS
lens (or its SIE equivalent)
\begin{equation} \label{E radius}
 \theta_E = 4 \pi
\frac{\sigma_{SIS}^2}{c^2} \frac{D_{ls}}{D_s}
\end{equation}
depends on the cosmological model through the ratio of
(angular-diameter) distances between lens and source and between
observer and lens. The angular diameter distance in flat
Friedmann-Robertson-Walker cosmology reads:
\begin{equation} \label{angular}
D(z;{\mathbf p}) = \frac{1}{1+z} \frac{c}{H_0}  \int_0^z
\frac{dz'}{h(z';{\mathbf p})}
\end{equation} where $H_0$ is the present value
of the Hubble function and $h(z;{\mathbf p})$ is a dimensionless
expansion rate dependent on redshift $z$ and cosmological model
parameters ${\mathbf p}$. For example in flat $\Lambda$CDM model
$h(z;{\mathbf p}) = \sqrt{\Omega_m (1+z)^3 + \Omega_{\Lambda}}$ we
have $\Omega_{\Lambda} = 1 - \Omega_m$ hence ${\mathbf p} = \{
\Omega_m \}$ in this case. Expansion rates in other cosmological
scenarios are given in Section 4. From now on we will assume
spatial flatness of the Universe since it is strongly supported by
independent and precise experiments, e.g. a combined WMAP5, BAO
and supernova data analysis gives $\Omega_{tot} =
1.0050^{+0.0060}_{-0.0061}$ (Hinshaw et al. 2009). The sample upon
which we work is small, and addition of (otherwise well
constrained) curvature parameter would only distort the results.

Provided one has reliable knowledge about the lensing system: i.e.
the Einstein radius $\theta_E$ (from image astrometry) and stellar
velocity dispersion $\sigma_{SIS}$ (form central velocity
dispersion $\sigma_0$ obtained from spectroscopy) one can use it
to test the background cosmology. This method is independent on
the Hubble constant value (which gets cancelled in the distance
ratio) and is not affected by dust absorption or source
evolutionary effects. It depends, however, on the reliability of
lens modelling (e.g. SIS or SIE assumption) and measurements of
$\sigma_0$. Hopefully, starting with the Lens Structure and
Dynamics (LSD) survey and the more recent SLACS survey
spectroscopic data for central parts of lens galaxies became
available allowing to assess their central velocity dispersions.
In practice central velocity dispersion $\sigma_0$ is estimated
from the velocity dispersion within $R_e / 8$ where $R_e$ is
optical effective radius. Thorough discussion of these issues can
be found in (Treu et al. 2006, Grillo st al. 2008) where the
arguments in favor of using $\sigma_0$ as representative to
$\sigma_{SIS}$ are presented. Moreover, there is a growing
evidence for homologous structure of late type galaxies (Treu et
al. 2006, Koopmans et al.2006, 2009) supporting reliability of
SIS/SIE assumption. In particular it was shown in (Koopmans et al.
2009) that inside one effective radius massive elliptical galaxies
are kinematically indistinguishable from an isothermal ellipsoid.

In the method used in this paper
%unlike the standard candles Hubble diagrams,
cosmological model enters not through a distance measure directly,
but rather through a distance ratio
\begin{equation} \label{observable}
{\cal D}^{th}(z_l,z_s; {\mathbf p}) = \frac{D_{s}({\mathbf
p})}{D_{ls}({\mathbf p})} = \frac{ \int_0^{z_s} \frac{dz'}{
h(z';{\mathbf p})}}{\int_{z_l}^{z_s} \frac{dz'}{ h(z';{\mathbf
p})}}
\end{equation}
and respective observable counterpart reads:
$$
{\cal D}^{obs} = \frac{4 \pi \sigma_0^2}{c^2 \theta_E}
$$
This has certain consequences both advantageous and
disadvantageous. The positive side is that the Hubble constant
$H_0$ gets cancelled, hence it does not introduce any uncertainty
to the results. On the other hand we have a disadvantage that the
power of estimating $\Omega_m$ is poor (which could be seen by
inspection into specific formulae for $h(z;{\mathbf p})$ -- see
Table 1 below). Therefore we only attempted to fit $\Omega_m$ in
the case of $\Lambda$CDM model (where it is the only free
parameter in flat cosmology) and it was successful only for the
restricted sample (see below). In other cases we assumed fixed
values for $\Omega_m$. Cosmological model parameters (coefficients
in the equation of state) have been estimated by minimizing the
chi-square:
\begin{equation} \label{chi2}
\chi^2({\mathbf p}) = \sum_i \frac{({\cal D}_i^{obs} - {\cal
D}_i^{th}({\mathbf p}))^2}{\sigma_{{\cal D},i}^2}
\end{equation}
where the sum is over the sample and $\sigma_{{\cal D},i}^2$
denotes the variance of ${\cal D}^{obs}$ (contextual use of the
same symbol for variances and velocity dispersions should not lead
to confusion). In calculating $\sigma_{{\cal D}}$ we assumed that
only velocity dispersion errors contribute and the Einstein radii
are determined accurately.

\section{Samples used}

We used a combined sample of $n=20$ strong lensing systems with
good spectroscopic measurements of central dispersions from the
SLACS and LSD surveys (essentially the same sample as used by
Grillo et al. (2008)). Original data concerning SLACS sample came
from Treu et al. (2006) (see also an erratum (Treu et al. 2006a) -
very important one). Data concerning LSD lenses are taken after
Treu and Koopmans (2004), Koopmans and Treu (2003, 2002).

As already noticed in Treu et al. (2006) the SLACS sample has an
average $D_{ls}/D_s$ ratio equal to 0.58 with an rms scatter 0.15.
Whereas for their purpose it was advantageous, in our context it
weakens the performance of the method. Therefore we selected a
sub-sample of $n=7$ lenses with the ${\cal D}$ ratio deviating
from the mean more than rms in either direction. It is summarized
in Table 1 where the names of lenses in the restricted sample are
given in bold.

\begin{table}
\caption{Combined SLACS + LSD lens sample. The restricted sample
(see text for explanation) is outlined in bold. }

\begin{tabular}{|c|c|c|c|c|} \hline
 Lens ID & $z_l$ & $z_s$ & $\theta_E ['']$ & $\sigma_0\;[km/s]$
  \\
 \hline

SDSS J0037-0942 & 0.1955 & 0.6322 & 1.47 & $ 282 \pm 11$ \\
 {\bf SDSS J0216-0813} & 0.3317 & 0.5235 & 1.15 & $ 349 \pm 24$ \\
 {\bf SDSS J0737+3216} & 0.3223 & 0.5812 & 1.03 & $ 326 \pm 16$ \\
SDSS J0912+0029 & 0.1642 & 0.3240 & 1.61 & $ 325 \pm 12$ \\
SDSS J0956+5100 & 0.2405 & 0.4700 & 1.32 & $ 318 \pm 17$ \\
 {\bf SDSS J0959+0410} & 0.1260 & 0.5349 & 1.00 & $ 229 \pm 13$ \\
SDSS J1250+0523 & 0.2318 & 0.7950 & 1.15 & $ 274 \pm 15$ \\
 {\bf SDSS J1330-0148} & 0.0808 & 0.7115 & 0.85 & $ 195 \pm 10$ \\
SDSS J1402+6321 & 0.2046 & 0.4814 & 1.39 & $ 290 \pm 16$ \\
 {\bf SDSS J1420+6019} & 0.0629 & 0.5352 & 1.04 & $ 206 \pm 5$ \\
SDSS J1627-0053 & 0.2076 & 0.5241 & 1.21 & $ 295 \pm 13$ \\
SDSS J1630+4520 & 0.2479 & 0.7933 & 1.81 & $ 279 \pm 17$ \\
SDSS J2300+0022 & 0.2285 & 0.4635 & 1.25 & $ 305 \pm 19$ \\
SDSS J2303+1422 & 0.1553 & 0.5170 & 1.64 & $ 271 \pm 16$ \\
 {\bf SDSS J2321-0939} & 0.0819 & 0.5324 & 1.57 & $ 245 \pm 7$ \\
{\bf Q0047-2808} & 0.485 & 3.595 & 1.34 & $ 229 \pm 15$ \\
CFRS03.1077 & 0.938 & 2.941 & 1.24 & $ 251 \pm 19$ \\
HST 14176 & 0.810 & 3.399 & 1.41 & $ 224 \pm 15$ \\
HST 15433 & 0.497 & 2.092 & 0.36 & $ 116 \pm 10$ \\
MG 2016& 1.004 & 3.263 & 1.56 & $ 328 \pm 32$ \\

\hline

\end{tabular}
\end{table}

For comparison of our results with the data which triggered the
dark energy problem, we also performed fits to the SNIa data
(n=307 supernovae) using Union08 compilation by Kowalski et al.
(2008). The $\Omega_m = 0.27$ prior was used throughout, except in
the  $\Lambda$CDM model where the fit was attempted.

\section{Cosmological models tested}

Several scenarios have been put forward as an explanation of
presently accelerating expansion of the Universe.
 The most obvious candidate is the cosmological constant
$\Lambda$ representing the energy of the vacuum. Corresponding
cosmological model, which turned out to be in agreement with all
existing (independent and alternative) observations is the
$\Lambda$CDM model. It is equivalent to $w = -1$ in the cosmic
equation of state $p= w \rho$ and the only free parameter here is
the $\Omega_m$ representing the density of baryonic plus cold dark
matter as a fraction of critical density (as already said spatial
flatness is assumed). On one hand it is therefore the most
parsimonious one, but well known fine tuning problems led many
people to seek beyond the $\Lambda$ framework and to develop the
concept of quintessence. Usually the quintessence is described in
a phenomenological manner, as a scalar
 field with an appropriate potential. % \cite{Ratra}.
 In first approximation it could be tested observationally
 by promoting $w$ to the role of a free parameter to be fitted from
 the data. However there is no a priori reason to expect that $w$
 should then be a constant.
 The parametrization of $w(z) = w_0 + w_a \frac{z}{1+z}$ developed by Chevalier \& Polarski (2001)
 and Linder (2003) turned out to be well suited and robust for such case.
 In the past, alternative parametrization $w(z) = w_0 + w_1 z$ was
 used (which is a truncated Taylor series representation of
 $w(z)$). Chevalier-Polarski-Linder parametrization instead uses an expansion with
 respect to the physical degree of freedom i.e. the scale factor
 (expanded around its present value). Dimensionless (i.e. with $H_0$ factored out) expansion
 rates for respective models are given in Table 2.

\begin{table*}
\begin{minipage}{110mm}
\caption{Equation of state and expansion rates $h(z)$ in the
models tested.} \label{h}
\begin{tabular}{@{}lll}

Model &EOS $p=w \rho$ &Cosmological expansion rate $h(z)$ \\
\hline

$\Lambda$CDM & $w = -1$ & $h(z) = \sqrt{ \Omega_m \; (1+z)^3 +
\Omega_{\Lambda}}$ \\

 Quintessence&$w = const.$&$h(z) = \sqrt{ \Omega_m \; (1+z)^3 + \Omega_Q \; (1+z)^{3(1+w)}} $\\

Chevalier-Polarski-Linder& $w(z) =  w_0 + w_a \frac{z}{1+z}$ &
$h(z) = \sqrt{ \Omega_m \; (1+z)^3 + \Omega_Q \;
(1+z)^{3(1+w_0+w_a)}\;\exp({-3 w_a z \over 1+z})} $\\
\hline
\end{tabular}
\end{minipage}
\end{table*}

For comparison we also performed fits of the models considered
above to the supernova Ia data with the same prior assumptions
(spatial flatness of the Universe and $\Omega_m$). We have taken
Union08 compilation (Kowalski et al. 2008) and instead of
straightforward $\chi^2$ fitting $m(z)$ vs. $D_L(z)$ (where $m$ is
visible magnitude $D_L$ denotes luminosity distance) we used a
well known modified approach equivalent to marginalizing over the
intercept (Nesseris \& Perivolaropoulos 2005).
\begin{figure}
  \vspace*{174pt}
  \caption{Best fits (dots) and (68 \%, 95\%) confidence regions for Chevalier-Linder-Polarski parameters
  in cosmic equation of state obtained from the full SLACS+LSD
  sample of lenses and Union08 SNIa data. A dot corresponding  to
  $\Lambda$CDM model is added for reference.}
\end{figure}

\begin{figure}
  \vspace*{174pt}
  \caption{Best fits (dots) and (68 \%, 95\%) confidence regions for Chevalier-Linder-Polarski parameters
  in cosmic equation of state obtained from restricted
  sample of lenses and Union08 SNIa data. A dot corresponding  to
  $\Lambda$CDM model is added for reference.}
%\centering
%\includegraphics{LinderRestrU08a.eps}
\end{figure}

\section{Results and conclusions}

Performing fits of different cosmological scenarios (shown in
Table 2) on the full SLACS + LSD sample of $n=20$ strong lensing
systems we obtained the equation of state parameters displayed in
Table 3. In $\Lambda$CDM model, %and braneworld scenario,
where $\Omega_m$ was the only free parameter we were not able to
make a reliable fit on the sample considered. As already mentioned
the reason of this is twofold. First, theoretical observable was
the ratio of two integrals differing only by the limits of
integration. Second, in the full sample the $D_{ls}/D_s$ ratio is
concentrated around a central value of $0.54$. Therefore, in case
of `simple' dependence (just a factor) of $h(z)$ on a parameter
(as is the case for $\Omega_m$) the bulk of the sample only
introduces an undesired scatter. More `sophisticated' dependence
(exponent of the integration variable)
 on the %$w$
equation of state parameters makes it possible to obtain fits on
this sample.

For comparison we also report (Table 4) values of these parameters
best fitted to the Union08 SNIa compilation (Kowalski et al.
2008). One can see that $w$ coefficient obtained from the full
strong lensing sample agrees with respective value derived from
supernovae data (almost whole $2\sigma$ confidence interval for
$w$ from Union08 data set lies within $1\sigma$ CI from lenses).
The value inferred is also in agreement with the WMAP5 results
presented in (Hinshaw et al. 2009) including also combined WMAP5,
BAO and SNIa analysis. Note that this is also consistent with the
$\Lambda$CDM model. Concerning evolving equation of state in
Chevalier-Polarski-Linder parametrization confidence regions in
the $(w_0,w_a)$ plane are shown in Figure 1. It can be seen that
the concordance model ($\Lambda$CDM) while consistent with SNIa
data (at $2\sigma$ level) is inconsistent with strong lensing
systems method applied here. SNIa results and strong lensing
results are marginally consistent with each other. Most probably
this is due to small sample of strong lenses combined with
systematics discussed above.

Working on the restricted sample (containing lenses with
 $D_{ls}/D_s$ ratio greater than rms spread around mean value)
 although decreased sample size dramatically (down to $n=7$)
 allowed to obtain fits on $\Omega_m$ in $\Lambda$CDM (Table 3) which turned
 out to agree with SNIa fits (Table 4) and WMAP5 data (Hinshaw et
 al. 2009). Although the best fit for $w$
 parameter quintessence scenario is higher than inferred from SNIa
 or WMAP5, yet the $2\sigma$ interval for Union08 data falls
 within $2\sigma$ interval from lenses. Hence the agreement is
 quite good. Similarly fits for $w_0$ and $w_a$ are improved
 (even though confidence regions get larger).

One should note however that systematic shift downwards in $(w_0,
w_a)$ plane persists. Such shift in best fitted parameters
inferred from supernovae (standard candles, sensitive to
luminosity distance) and BAO or acoustic peaks (standard rulers,
sensitive to angular diameter distance) has already been noticed
and discussed by Lazkoz, Nesseris \& Perivolaropoulos (2007) and
by Linder \& Roberts (2008). Bearing in mind similar mutual
inconsistency in the Hubble constant values inferred from lensing
time delays $H_0 = 52 \pm 6 \; km\; s^{-1}\;Mpc^{-1}$ (Kochanek \&
Schechter 2003) and from the HST Key Project $H_0 = 72 \pm 8 \;
km\; s^{-1}\;Mpc^{-1}$ (Freedman et al. 2001) our result suggests
the need for taking a closer look at compatibility of results
derived by using angular diameter distances and luminosity
distances respectively. It is also worth noticing that the ideas
of testing the Etherington reciprocity relation between these two
distance measures have been discussed by Basset \& Kuntz (2004)
and by Uzan, Aghanim \& Mellier (2005).

In conclusion our results demonstrated that the method discussed
in Biesiada (2006) and extensively investigated by Grillo et al.
(2008) on simulated data can be used in practice to constrain
cosmological models. It turned out to give reasonable results on
already accessible samples of strongly lensed systems. Besides the
uncertainties related to velocity dispersion measurements and
their conversion to relevant lens model parameters (as well as the
impact of SIS assumption) the issue of systematics associated with
$D_{ls}/D_s$ ratio in the sample turned out to be important. In
particular it implies that strong lensing survey strategies like
the one adopted in SLACS survey are better from this point of
view. Lensing systems are gathered around something like $0.58$ in
distance ratio because it is roughly the configuration for which
lensing probability (for a given lens mass) is the highest.
Earlier searches were focused on source population (quasars)
seeking for close pairs or multiples and checking if they are
multiple images of a single source lensed by an intervening
galaxy. Therefore a high lensing probability was an important
selection factor there. On the other hand SLACS survey is focused
on possible lens population (massive elipticals) with good
spectroscopic data. Using SDSS templates spectra are carefully
checked for residual emission (at least three distinct common
atomic transitions) coming from higher redshifts. Such candidates
undergo image processing by subtracting parametrized brightness
distribution typical for early type galaxies in order to reveal
multiple images of the quasar. Details can be found in Bolton et
al. (2006). Therefore, besides the obvious bonus of having central
velocity dispersion measured, such strategy is better suited for
discovering systems with larger $D_{ls}/D_s$ ratios which in turn
can be used for testing cosmological models.

Finally, one important effect -- neglected here -- should be
mentioned, which is the influence of line of sight mass
contamination. The debate on this issue started with Bar-Kana
(1996) and Keeton et al. (1997) who were among the first to
convincingly demonstrate that the effect of large scale structure
on strong lensing could be significant. More recent results on
this issue can be found in Dalal et al. (2005) (in the context of
cluster lensing) or Momcheva et al.(2006). This rises the issue of
an impact this effect might have on our results, since the sample
was small. Straightforward naive first guess (based on Poissonian
statistics) might suggest that sample size of order of a few
hundred lenses might reduce line of sight `noise' contamination
down to a few percent. This is however not that simple since the
line of sight contamination is in fact a systematic effect.
Namely, massive early type galaxies (ie. typical lenses) prefer
overdense environments, so one consistent approach would be to
follow light rays (ray-shooting simulation) through many lens
planes (obtained from cosmological N-body simulation) up to high
source redshift. This was done by Wambsganss et al. (2005) with
the result that up to $z_s=1$ most (i.e. $95\%$ ) of lenses
involved only a single mass concentration, whereas for sources at
$z_s=3.8$ important contribution of intervening mass could be
significant in $38\%$ of strong lensing systems. This result
suggests that the line of sight contamination should be addressed
separately for each particular survey. For the SLACS survey (where
the bulk of our sample comes from) this was assessed in Treu et
al. (2009) where the authors found that SLACS lens galaxies are
unbiased population (i.e. with environmental effects typical to
the over-all population of early type galaxies) and typical
contribution from external mass distribution is small -- no more
than a few percent. Fortunately, the SLACS survey is ongoing i.e.
the sample of spectroscopically investigated strong lenses is
growing. However this survey is relatively shallow, so for
cosmological applications one is forced to combine it with deeper
surveys (with different designs -- hence different systematics)
and the problem of line of sight contamination remains
challenging.

\begin{table}
\caption{Fits to different cosmological models from SLACS + LSD
lens samples ($n=15 + 5 = 20$). Fixed value of $\Omega_m =0.27$
was assumed.}

\begin{tabular}{|c|c|c|}
\hline
 Cosmological model & Best fit parameters (with $1\sigma$) &
$\chi^2 /dof$  \\
 \hline

$\Lambda$CDM & not possible & \\
Quintessence & $w = -0.9829 \pm 0.2415$ & %$\chi^2 = 64.834\; \chi^2/dof =
$3.41$\\
Chevalier-Linder-Polarski & $w_0 = 1.2605 \pm 0.8177 $ & %$\chi^2 = 54.948\; \chi^2/dof =
$3.05$\\
    & $ w_a = -9.4443 \pm 4.4193 $ & \\
 \hline

\end{tabular}
\end{table}

\begin{table}
\caption{Fits to different cosmological models from restricted
SLACS + LSD lens sample ($=7$). Fixed value of $\Omega_m =0.27$
assumed, besides $\Lambda$CDM , where the fit was successful.}

\begin{tabular}{|c|c|c|}
\hline
 Cosmological model & Best fit parameters (with $1\sigma$) & $\chi^2/dof$  \\
 \hline

$\Lambda$CDM & $\Omega_m = 0.2660 \pm 0.2796$ & %$\chi^2 = 10.5531\; \chi^2/dof =
$1.76 $\\
Quintessence & $w = -0.6320 \pm 0.4461  $ & %$\chi^2 = 23.457\; \chi^2/dof =
$3.91 $ \\
Chevalier-Linder-Polarski & $w_0 = 0.3588 \pm 1.2453  $ & %$\chi^2 = 9.4054\; \chi^2/dof =
$1.88$ \\
        &  $ w_a = -3.6301 \pm 5.3278 $ & \\

\hline

\end{tabular}
\end{table}

\begin{table}
\caption{Fits to different cosmological models from Union08 sample
$n=307$ SNIa. Prior $\Omega_m =0.27$ assumed.}

\begin{tabular}{|c|c|c|}
\hline
 Cosmological model & Best fit parameters (with $1\sigma$) &
$\chi^2 /dof$  \\
 \hline

$\Lambda$CDM & $\Omega_m = 0.287 \pm 0.027$ & %$\chi^2 = 311.936\;\chi^2/dof =
$1.02$\\
Quintessence & $w = -1.061 \pm 0.083$ & %$\chi^2 = 311.583\; \chi^2/dof =
$1.02$\\
Chevalier-Linder-Polarski & $w_0 = -1.263 \pm 0.257 $ & %$\chi^2 = 310.911\; \chi^2/dof =
$1.02$\\
            & $ w_a = 1.254 \pm 1.484 $ & \\

\hline

\end{tabular}
\end{table}

\section*{Acknowledgements}
Authors thank to the referee for suggesting inclusion of line of
sight contamination effects, which was very beneficial for the
paper. The work was supported by the Polish Ministry of Science
Grant no N N203 390034


\begin{thebibliography}{99}

\bibitem{Alcaniz}
Alcaniz J. S., Jain D., Dev A., 2003, Phys. Rev. D, 67, 043514

\bibitem{Allen}
Allen S. W., et al. 2008, MNRAS, 383, 879

\bibitem{Bar-Kana}
Bar-Kana R., 1996, ApJ, 468, 17

\bibitem{Basset}
Bassett B. A., Kunz M., 2004, Phys. Rev. D 69, 101305

\bibitem{Biesiada2006}
Biesiada M., 2006, Phys. Rev. D,  73, 023006

\bibitem{Bolton}
Bolton A.S., et al., 2006, ApJ, 638, 703

\bibitem{Buchert}
Buchert T., 2001, Gen. Relativ. Grav., 33, 1381

\bibitem{Chevalier}
Chevalier M., Polarski D., 2001, Int. J. Mod. Phys. D, 10, 213

\bibitem{Dalal}
Dalal N., Hennavi J.F., Bode P., 2005, ApJ, 622, 99

\bibitem{Davis}
Davis T. M., et al. 2007, ApJ 666, 716

\bibitem{Eisenstein}
Eisenstein D. J., et al. 2005, ApJ, 633, 560

\bibitem{Ellis}
Ellis G.F.R. \& Buchert T., 2005, Phys. Lett. A, 347, 38

\bibitem{Freedman}
Freedman W.L., et al., 2001,
%Final results from the HST Key Project to measure the Hubble Constant,
ApJ, 553, 47

\bibitem{Futamase}
Futamase T., Yoshida S., 2001, Prog. Theor. Phys., 105, 887

\bibitem{Grillo}
Grillo C., Lombardi M., Bertin G., 2008, Astron.Astrophys., 477,
397

\bibitem{WMAP5}
Hinshaw G. et al., 2009, ApJ.Suppl., 180, 225

\bibitem{Keeton}
Keeton C.R., Kochanek C.S., Seljak U., 1997, ApJ, 482, 604

\bibitem{Kochanek SAAS}
Kochanek C.S. and Schechter P.L., The Hubble constant from
gravitational lens time delays, Carnegie Observatories
Astronomical Series, vol.2: Measuring and Modelling the Universe
ed. W.L.Frreedman (Cambridge: Cambridge University Press) 2003
arXiv:astro-ph/0306040

\bibitem{Koopmans02}
Koopmans L.V.E, Treu T., 2002, ApJ, 568, L5

\bibitem{Koopmans03}
Koopmans L.V.E., Treu T., 2003, ApJ, 583, 606

\bibitem{Koopmans06}
Koopmans L.V.E., et al., 2006, ApJ, 649, 599

\bibitem{Koopmans09}
Koopmans L.V.E., et al., 2009, arXiv:0906.1349

\bibitem{Kowalski}
Kowalski M., et al., 2008, ApJ, 686, 749

\bibitem{Lazkoz}
Lazkoz R., Nesseris S., Perivolaropoulos L., 2008, JCAP 0807, 012

\bibitem{Linder}
Linder E. V., 2003, Phys. Rev. D, 68, 083503

\bibitem{LinderRoberts}
Linder E.V., Roberts G.,

\bibitem{Momcheva}
Momcheva I., Williams K., Keeton C.R., Zabludoff A., 2006, ApJ,
641, 169

\bibitem{Nesseris}
Nesseris S., Perivolaropoulos L., 2005, Phys. Rev. D 72, 123519

\bibitem{Perlmutter}
Perlmutter, S., et al., 1999, ApJ, 517, 565

\bibitem{Rasanen}
R{\"a}s{\"a}nen S., 2004, JCAP 02,  003

\bibitem{Riess98}
Riess, A. G., et al., 1998, AJ, 116, 1009

\bibitem{Riess04}
Riess, A. G., et al., 2004, ApJ, 607, 665

\bibitem{Falco}
P. Schneider, J. Ehlers, E. E. Falco, Gravitational Lenses
(Springer Verlag, Berlin, 1992).

\bibitem{Silva}
Silva P.T., Bertolami O., 2003, ApJ, 599, 829

\bibitem{Spergel}
Spergel D. N., et al., 2003, ApJS, 148, 175

\bibitem{Treu04}
Treu T., Koopmans L.V.E., 2004, ApJ, 611, 739

\bibitem{Treu06a}
Treu T. et al., 2006a, ApJ, 640, 662

\bibitem{Treu06b}
Treu T., et al., 2006b, ApJ, 650, 1219

\bibitem{Treu09}
Treu et al. 2009 ApJ, 690, 670

\bibitem{Uzan}
Uzan J.P., Aghanim N., Mellier Y., 2004, Phys. Rev. D, 70, 083533

\bibitem{Wambsganss}
Wambsganss J., Bode P., Ostriker J.P., 2005, ApJ, 635, L1

\bibitem{Wiltshire}
Wiltshire D., 2007, Phys. Rev. Lett., 99, 251101

\bibitem{Wood-Wasey}
Wood-Vasey W. M., et al., 2007, ApJ, 666, 694




\end{thebibliography}
\end{document}